\documentclass{aa}
\usepackage{graphicx,txfonts,amssymb}

\def\mes{M{\'e}sz{\'a}ros}
\def\etal{et al.\ }
\def\HETE{\mbox{HETE--2}}
\def\Chandra{\textit{Chandra}}
\def\AV{0.42}
\def\tb{t_\mathrm{b}}
\def\NH{N_\mathrm{H}}
\def\Xray{\mbox{X--ray}}
\def\nui{\nu_\mathrm{i}}\def\nuc{\nu_\mathrm{c}}\def\nuo{\nu_\mathrm{o}}

\begin{document}

\title{Optical and NIR Observations of the Afterglow of GRB\,020813%
\thanks{Based on observations partly made with ESO telescopes at the Paranal Observatories under programme Id 69.D-0461
and with the Italian TNG telescope under programme TAC 8\_01(47).}}
\titlerunning{Lightcurve of GRB\,020813}

\author{S.~Covino\inst{1} \and D.~Malesani\inst{1} \and F.~Tavecchio\inst{1} \and
L.A.~Antonelli\inst{2} \and A.~Arkharov\inst{3} \and A.~Di~Paola\inst{2} \and D.~Fugazza\inst{4} \and G.~Ghisellini\inst{1}
\and  V.~Larionov\inst{5,6} \and D.~Lazzati\inst{7} \and F.~Mannucci\inst{8} \and N.~Masetti\inst{9} \and
R.~Barrena\inst{4} \and S.~Benetti\inst{10} \and A.J.~Castro--Tirado\inst{11} \and S.~Di~Serego~Alighieri\inst{12} \and
F.~Fiore\inst{2} \and F.~Frontera\inst{9,13} \and A.~Fruchter\inst{14} \and F.~Ghinassi\inst{4} \and M.~Gladders\inst{15}
\and P.B.~Hall\inst{16,17} \and G.L.~Israel\inst{2} \and S.~Klose\inst{18} \and A.~Magazz\`u\inst{4}
\and E.~Palazzi\inst{9} \and M.~Pedani\inst{4} \and E.~Pian\inst{19} \and P.~Romano\inst{1} \and M.~Stefanon\inst{1} \and
L.~Stella\inst{2}}

\authorrunning{Covino \etal}
\offprints{S.~Covino \email{covino@mi.astro.it}}

\institute{  
INAF, Osservatorio Astronomico di Brera, via E. Bianchi 46, 23807 Merate (LC), Italy.
\and         
INAF, Osservatorio Astronomico di Roma, via Frascati 33, Monteporzio Catone (Roma), Italy.
\and         
Central Astronomical Observatory at Pulkovo, Pulkovskoe shosse 65, 196140 Saint Petersburg, Russia.
\and         
INAF, Telescopio Nazionale Galileo, Roque de Los Muchachos, PO box 565, 38700 Santa Cruz de La Palma, Spain.
\and         
St. Petersburg University, St. Petersburg, Petrodvorets, Universitetsky pr. 28, 198504 St. Petersburg, Russia.
\and         
Isaac Newton Institute of Chile, St. Petersburg Branch.
\and         
Institute of Astronomy, University of Cambridge, Madingley Road, CB3 0HA Cambridge, UK.
\and         
Istituto di Radioastronomia, CNR, largo E. Fermi 5, 50125 Firenze, Italy.
\and         
Istituto di Astrofisica Spaziale e Fisica Cosmica, via Gobetti 101, 40129 Bologna, Italy.
\and        
INAF, Osservatorio Astronomico di Padova, vicolo dell'Osservatorio 5, 35122 Padova, Italy.
\and        
Instituto de Astrof{\'\i}sica de Andaluc{\'\i}a, CSIC, P.O. Box 03004, E-18080 Granada, Spain.
\and        
INAF, Osservatorio Astrofisico di Arcetri, largo E. Fermi 5, 50125 Firenze, Italy.
\and        
Dipartimento di Fisica, Universit\`a di Ferrara, Via Paradiso 12, 44100 Ferrara, Italy.
\and        
Space Telescope Science Institute, 3700 San Martin Drive, Baltimore, MD 21218, USA.
\and        
Carnegie Observatories, 813 Santa Barbara Street, Pasadena, CA 91101-1292.
\and        
Departamento de Astronom\'{\i}a y Astrof\'{\i}sica, PUCA, Casilla 306, Santiago 22, Chile.
\and        
Princeton University Observatory, Princeton, NJ 08544-1001, USA.
\and        
Th\"uringer Landessternwarte Tautenburg, Karl-Schwarzschild-Observatorium, Sternwarte 5, 07778 Tautenburg, Germany.
\and        
INAF, Osservatorio Astronomico di Trieste, via Tiepolo 11, 34131 Trieste, Italy.
}

\date{Received }

\abstract{
We present optical and near-infrared (NIR) photometry of the bright afterglow of GRB\,020813. Our data span from
$3$~hours to $4$~days after the GRB event. A rather sharp achromatic break is present in the light curve, $14$~hours
after the trigger. In the framework of jetted fireballs, this break corresponds to a jet half-opening angle
of~$1.9\degr\pm0.2\degr$, the smallest value ever inferred for a GRB. We discuss our results in the framework of
currently available models, and find that they have problems in explaining the joint temporal and spectral properties,
and in particular the slow decay before the break.
\keywords{gamma rays: bursts -- radiation mechanisms: non-thermal}
}

\maketitle

\section{Introduction}\label{sec:intro}

Since the discovery of the first afterglow of a gamma-ray burst (GRB; Costa \etal \cite{Co97}; van Paradjis \etal
\cite{vP97}), our knowledge of these mysterious explosions has rapidly increased. The huge energetics implied by their
cosmological distance (e.g. Metzger \etal \cite{Me97}) has severely constrained existing models, leading to the standard
internal/external shock scenario (e.g. Piran \cite{Pi99}; \mes{} \cite{Me02} and references therein). According to
this model, the prompt gamma-ray emission is produced by internal collisions in a relativistic blastwave, while the
afterglow originates by the slowing down of the same fireball in the surrounding medium.

An important question still remains unsettled: are the fireballs spheres or jets? In the latter case, a steepening in
the light curve is expected when the bulk Lorentz factor of the fireball equals the inverse of the opening angle of the
jet (Rhoads \cite{Rh99}). This steepening must occur at all frequencies and at the same time (`achromatic' break). To
date, there are several possible examples of such behaviour in the optical band. The most convincing cases are perhaps
GRB\,990510 (e.g. Israel \etal \cite{Is99}; Harrison \etal \cite{Ha99}; Stanek \etal \cite{St99}) and GRB\,010222
(e.g. Masetti \etal \cite{Ma01}; Sagar \etal \cite{Sa01}; Stanek \etal \cite{St01}), for which the break times were
$\sim 1.5$~days and $\sim 0.7$~days respectively.

GRB\,020813 was detected on 2002 August 13 by the \HETE{} spacecraft at 2:44:19 UT (Villasenor \etal \cite{Vi02}). It
was a bright event, lasting more than $125$~s, with a fluence of $\sim 3.8 \times 10^{-5}$~erg~cm$^{-2}$
($25\div100$~keV) as reported by Ulysses (Hurley \etal \cite{Hu02}). The rapid coordinate dissemination and the small
\HETE{} error box ($4'$ radius) allowed Fox, Blake \& Price (\cite{FBP02}) and Gladders \& Hall (\cite{GH02a}) to
independently identify a bright optical counterpart with position $\alpha_{2000} = 19^{\rm h}46^{\rm m}41\fs9$,
$\delta_{2000} = -19\degr36\arcmin04\farcs8$, just 1.9~hours after the trigger. The spectroscopic redshift, determined
with KECK\,1--LRIS, was $z = 1.255$ (Barth \etal \cite{Ba03}). Subsequent observations by various groups
allowed a preliminary sampling of the light curve, which exhibited a steepening in the optical bands beginning a few
hours after the GRB trigger (e.g. Bloom, Fox \& Hunt \cite{BFH02}).  Polarimetric observations were carried out by Barth
\etal (\cite{Ba03}) and Covino \etal (\cite{Co02}), yielding positive polarization signals at the level of $P =
(2.9\pm0.1)\%$ and $P = (1.17\pm0.16)\%$, $4.7$~hours and $23$~hours after the burst, respectively (both measurements
are uncorrected for the small Galactic-induced polarization). The afterglow was also observed in the \Xray{} band
($0.6\div6$~keV) by \Chandra, as a bright fading source with an average flux $F \sim
1.9\times10^{-12}$~erg~cm$^{-2}$~s$^{-1}$ (Butler \etal \cite{Bu03}). A positive radio detection was obtained by
Frail \& Berger (\cite{FB02}), with $F_\nu = 300$~$\mu$Jy at $8.46$~GHz.  Only upper limits were reported in the
millimeter region (Bremer \& Castro--Tirado \cite{BCT02}; Bertoldi \etal \cite{Be02}).

\section{Data collection, reduction, and analysis}\label{sec:data}

\begin{figure}
  \includegraphics[width=\columnwidth,keepaspectratio]{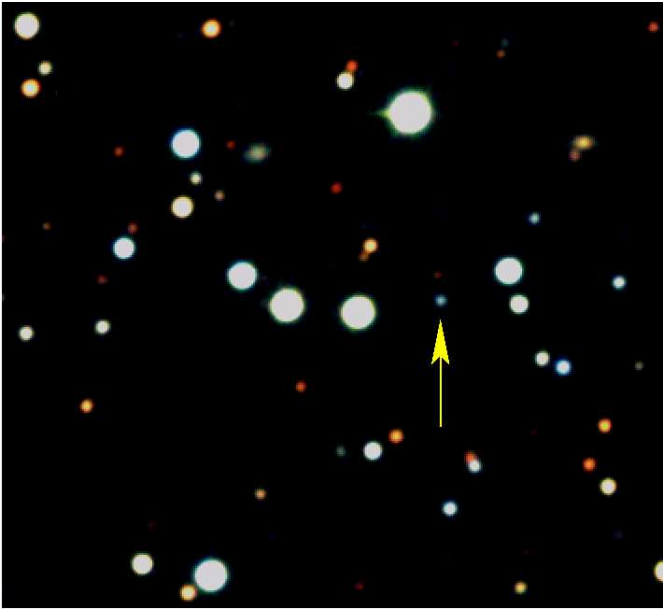}
  \caption{The field of GRB\,020813 ($\sim1.5\arcmin\times1.5\arcmin$), as imaged by TNG--DOLORES
  on Aug~15 UT. The afterglow is marked by an arrow. This is a composite image from $B$, $V,$ and $R$ filters. North is up, East is
  left.\label{fg:OT}}
\end{figure}

\begin{figure}[t]
\includegraphics[width=\columnwidth]{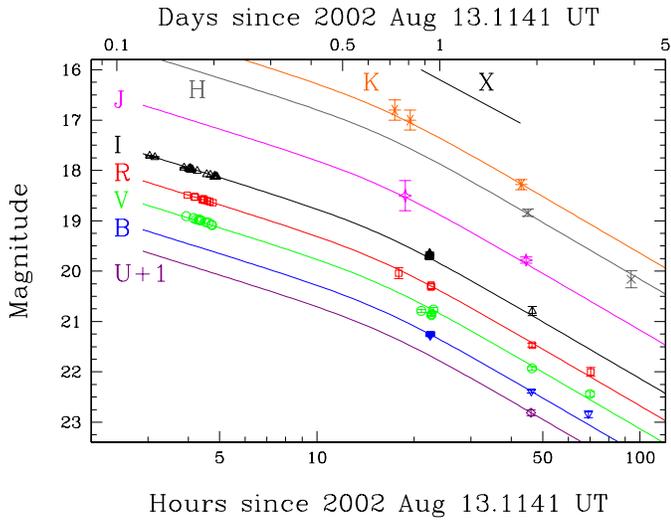}
\caption{Optical and NIR light curves of GRB\,020813 afterglow, with our best fit according to Eq.~\ref{eq:lc} (dashed
lines); refer to Table~\ref{tb:tfit} for fit parameters, and to the web page\protect\footnotemark[3] for the numerical
values. For the first night ($t<5$~h), error bars are smaller than symbols and are not plotted. The thick solid line
shows the (arbitrarily normalized) \Xray{} decay.}\label{fg:lc}
\end{figure}

We began observing the optical afterglow of GRB\,020813 (Fig.~\ref{fg:OT}) on Aug~13.99 UT, with the ESO
\mbox{VLT--UT3} (Melipal), equipped with FORS\,1 and a $V$ filter in imaging polarimetric mode; during the subsequent nights
(Aug 15, 16, and 17), we monitored the optical/NIR light curve using the 3.6~m Telescopio Nazionale Galileo (TNG) in the
Canary Islands. $UBVRI$ and $JHK$ data were acquired with the \mbox{DOLORES} and \mbox{NICS} instruments
respectively. Additional NIR images ($J$ and $K$) were taken with the AZT24 1.1~m telescope in Campo Imperatore
(L'Aquila, Italy) on Aug~13. We then added to our photometric set the publicly avaliable data%
\footnote{\texttt{ftp://ftp.ociw.edu/pub/gladders/GRB/GRB020813}\,.}
of Mike Gladders and Pat Hall (\cite{GH02b}), acquired with the Baade $6.5$~m (Magellan~1) telescope. Standard stars
were imaged during the night of Aug~16 with the TNG. Data reduction and calibration were carried out following standard
procedures, as implemented in the Eclipse package (version 4.2.1; Devillard \cite{De97}), while photometry was performed
by means of GAIA%
\footnote{\texttt{http://star-www.dur.ac.uk/$\sim$pdraper/gaia/gaia.html}\,.}.
All photometric data, shown in Fig.~\ref{fg:lc}, are posted online%
\footnote{\texttt{http://www.merate.mi.astro.it/$\sim$malesani/GRB/020813}\,.}.


We modelled the afterglow light curve with the functional form first proposed by Beuermann \etal (\cite{Be99}):
\begin{equation}\label{eq:lc}
F_\nu(t) = \frac{F_\nu^*}{\left[(t/\tb)^{k\delta_1} +(t/\tb)^{k\delta_2}\right]^{1/k}};
\end{equation}
here $F_\nu$ is the (monochromatic) flux at frequency $\nu$, and $t$ is the time elapsed from the GRB; this expression
reduces to $F_\nu \propto t^{-\delta_1}$ for $t \ll \tb$ and $F_\nu \propto t^{-\delta_2}$ for $t \gg \tb$; $\tb$ is the
break time at which the transition between the two regimes occurs, and $k>0$ describes how fast this transition takes
place (the larger $k$ the sharper the transition); $F_\nu^*$ is a normalization constant. We started by fitting the $V$,
$R$ and $I$ bands, which are sampled closely enough to constrain $\tb$ and $\delta_1$. To within the errors, the
inferred parameters, summarized in Table~\ref{tb:tfit}, were determined to be the same between the different bands, even
if there is a marginal hint (not statistically significant) for a correlation between $\tb$ and the frequency. We
caution that $\delta_2$ could also be larger, since our sampling does not extend very much after $\tb$. By fitting
simultaneously the whole dataset, including $U$, $B$ and NIR data, we got our best values. As shown in Fig.~\ref{fg:lc},
at early times the decay was fairly slow, similar to that seen in other bursts with an early break (e.g. GRB\,010222:
Masetti \etal \cite{Ma01}).  After $\tb$, the slope increased by $0.65\pm0.05$. In all fits, the parameter $k$, though
poorly constrained, was always $\gtrsim 5$, indicating a very sharp transition. This is confirmed by the fact that early
data ($t < 5$~h) are well fitted by a single powerlaw, showing that the steepening had not yet started at this
time. This is in contrast with the finding of Gladders \& Hall (\cite{GH02c}), who claim a steepening in the $I$-band
between $3$ and $5$ hours after the GRB.

\begin{table}
\caption{Fitted parameters describing the temporal evolution of the afterglow. $\delta_1$ and $\delta_2$ are the early- and
late-time temporal slopes, and $\tb$ is the break time (measured in hours since Aug 13.1141~UT); see
Eq.~\ref{eq:lc}. Errors are 1-$\sigma$.}
\centering\begin{tabular}{|llll|}\hline
  Band  &$\delta_1$     &$\delta_2$     &$\tb$ (h)    \\ \hline 
  $V$   &$0.84\pm0.11$  &$1.34\pm0.06$  &$11.9\pm6.4$ \\        
  $R$   &$0.76\pm0.05$  &$1.46\pm0.04$  &$13.7\pm1.2$ \\        
  $I$   &$0.80\pm0.03$  &$1.43\pm0.12$  &$16.7\pm1.7$ \\ \hline 
  All   &$0.78\pm0.02$  &$1.44\pm0.04$  &$14.2\pm0.8$ \\ \hline 
\end{tabular}\label{tb:tfit}
\end{table}

\begin{figure}
  \includegraphics[width=\columnwidth,keepaspectratio]{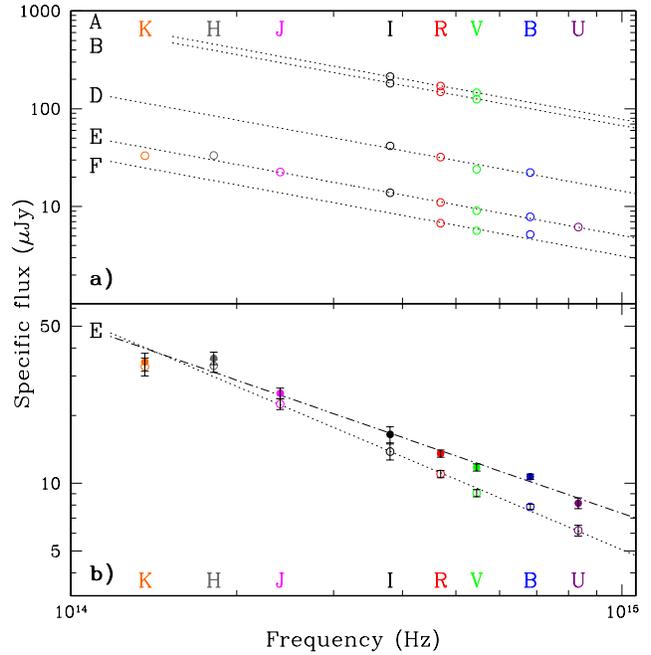}
  \caption{\textbf{a)} Photometric spectra of GRB\, 028013 afterglow at various times (see Table~\ref{tb:sfit}),
  together with the best fit (dotted lines). Data are corrected for reddening in the Galaxy but not in the
  host. Error bars are smaller than the symbols and are not plotted. \textbf{b)} TNG spectrum on Aug 15.04~UT; open
  points and dotted line are corrected only for Galactic reddening, while filled points and dot-dashed line are
  corrected also for reddening in the host (see text).\label{fg:spectrum}}
\end{figure}

To further check if the break is really achromatic, we studied the spectral properties of the afterglow, using our
photometric multiband data points, both before and after the break (Fig.~\ref{fg:spectrum}).  We first corrected for
Galactic reddening, assuming a neutral hydrogen column density%
\footnote{\texttt{http://heasarc.gsfc.nasa.gov/docs/corp/tools.html}\,.}
of $\NH = 7.52\times10^{20}$~cm$^{-2}$, corresponding to a $V$-band absorption $A_V = 0.42$ (Predehl \& Schmitt
\cite{PS95}). Individual powerlaw fits to each dataset yield the spectral indices reported in Table~\ref{tb:sfit}
($F_\nu \propto \nu^{-\alpha}$), consistent with $\alpha$ remaining constant during all the observations.  This is a
confirmation that the break is indeed achromatic. A simultaneous fit to the full dataset (six spectra) gives the best
value $\alpha = 1.04 \pm 0.03$. The colors are $B-V = 0.32\pm0.05$, $V-R = 0.39\pm0.06$, $R-I = 0.45\pm0.10$, within the
range of the sample by \v{S}imon \etal (\cite{Si01}). The chi-square of the fit is $\chi^2 = 82$ for 17 degrees of
freedom. Such large value is due to the use of data coming from different telescopes.

\begin{table}
\caption{Time evolution of the spectral index ($F_\nu \propto \nu^{-\alpha}$). Galactic (but not intrinsic)
extinction has been accounted for by assuming $A_V = \AV$.}
\centering\begin{tabular}{|lllll|} \hline
  UT    & \# & Telescope  & Bands      & $\alpha$      \\ \hline 
  13.28 & A  & Baade 6.5m & $VRI$      & $1.06\pm0.06$ \\        
  13.31 & B  & Baade 6.5m & $VRI$      & $1.00\pm0.05$ \\        
  13.88 & C  & AZT 1.1m   & $RJK$      & $0.85\pm0.14$ \\        
  14.04 & D  & Baade 6.5m & $BVRI$     & $1.15\pm0.05$ \\        
  15.04 & E  & TNG 3.5m   & $UBVRIJHK$ & $1.01\pm0.04$ \\        
  16.01 & F  & TNG 3.5m   & $BVR$      & $0.72\pm0.22$ \\ \hline 
  All   &--- & ---        & ---        & $1.04\pm0.03$ \\ \hline 
\end{tabular}\label{tb:sfit}
\end{table}

\section{Discussion}

The afterglow of GRB\,020813 represents a new case in which the optical light curve shows an achromatic break.  The
break was quite sharp and occurred early, similar to that of GRB\,010222. The slope after the break ($\sim1.4$) was
typical of afterglows at times between $1$ and $10$~days after the GRB. Since observations are often performed during
this range of days (and only rarely earlier), it is possible that many such breaks have been missed in the past. An
intriguing example of this situation is GRB\,010921, whose light curve might have displayed {\em two} breaks, the first
only constrained by an early LOTIS upper limit to be at less than $\sim1$~day (Park \etal \cite{Pa02}), and the second
determined by means of HST observation after $\sim 35$~days (Price \etal \cite{Pr03}).

There are various interpretations to explain the presence of an achromatic break. If the blastwave producing the GRB is
collimated, the break is expected to occur when the inverse of the bulk Lorentz factor of the ejecta $\Gamma$ equals the
half-opening angle of the jet $\vartheta_\mathrm{jet}$ (Rhoads \cite{Rh99}). In this context, Frail \etal (\cite{Fr01})
found that the net energy released by GRBs, after correcting for beaming, is approximately constant for a number of
events, to within a factor of $\sim2$. This result is based on the observed correlation between break times and measured
isotropic energies. Using the fluence $S = 3.8 \times 10^{-5}$~erg~cm$^{-2}$ reported by Hurley \etal (\cite{Hu02}) in
the $(25\div100)$~keV band, and applying a bolometric correction of $5.5\pm3.5$ (calculated in the same way as in Bloom,
Frail \& Sari \cite{BFS01}), we get an isotropic energy%
\footnote{We adopt a standard cosmology: $h_0 = 0.65$, $\Omega_\mathrm{m} = 0.3$, $\Omega_\Lambda = 0.7$.}
of $(9.8\pm6.2) \times 10^{53}$~erg, ranking third in the sample of bursts with known redshift after GRB\,990123 and
GRB\,000131. Using the formalism described by Frail \etal (\cite{Fr01}), the opening angle of the jet is
$\vartheta_\mathrm{jet} = (1.7\degr\pm0.2\degr) n_{-1}^{1/8}\eta_{20}^{1/8}$, where $n_{-1}$ is the external number
density in units of $10^{-1}$~cm$^{-3}$ and $\eta = 20\%\,\eta_{20}$ is the gamma-ray production efficiency. This
is the smallest value ever reported for a GRB. The beaming-corrected energy is hence $(5.4\pm2.5) \times 10^{50}
n_{-1}^{1/4}\eta_{20}^{1/4}$~erg, close to the value found by Frail \etal (\cite{Fr01}) in their sample.

\begin{figure}[t]
\includegraphics[width=\columnwidth,keepaspectratio]{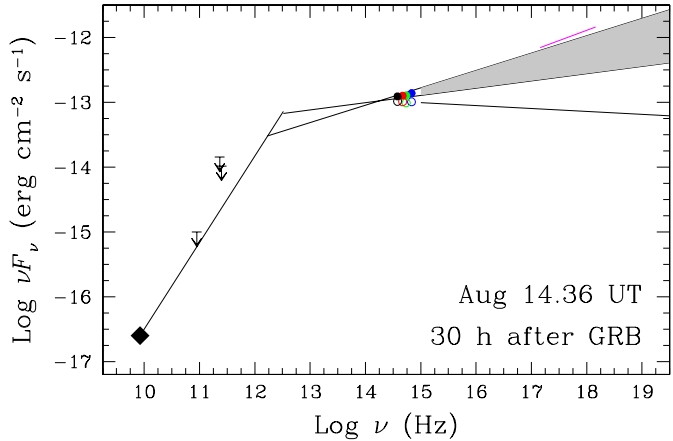}
\caption{Radio to \Xray{} spectral energy distribution of the afterglow of GRB\,020813, $30$~h after the
trigger. Optical ($BVRI$ bands) and \Xray{} data have been computed at the epoch of the radio detection by interpolating
between different spectra and using the appropriate decay law. Open symbols are corrected only for Galactic extinction,
while the additional host component has been taken into account for filled symbols. Radio detection together with the
millimeter upper bounds limit the spectral slope ($F_\nu \propto \nu^{-\alpha}$) to be $\alpha \ge -1/3$ at GHz
frequencies (in particular a self-absorbed spectrum is ruled out). In the region between the millimeter and optical
bands, the drawing is only indicative.}\label{fg:SED}
\end{figure}

In a different interpretation, an early break can be caused by the onset of the nonrelativistic phase, when $\Gamma$
drops to unity (e.g. Dai \& Lu \cite{DL99}). This interpretation has been proposed for GRB\,010222, which shows spectral
and temporal properties similar to the case presented here (Masetti \etal \cite{Ma01}). In order to stop the fireball so
quickly, however, an extremely high external density $n \sim 10^{8}$~cm$^{-3}$ is  required (Panaitescu \& Kumar
\cite{PK00}). Since the radius of the fireball at $t = \tb$ is $R \sim 10^{16}$~cm, the surrounding material must in
this case have a hydrogen column density of $\NH \sim nR \sim 10^{24}$~cm$^{-2}$, and hence it must be completely
ionized in order not to conflict with the \Xray{} observation, which does not show a high value of $\NH$ at the redshift
of the host (Butler \etal \cite{Bu03}). These parameters are typical of a supernova remnant (e.g. Vietri \& Stella
\cite{VS98}). If this interpretation is correct, however, the radio spectrum is expected to be heavily
self-absorbed; this is not the case, as can be seen in Fig.~\ref{fg:SED}.

\begin{table}
\caption{Predicted slopes and comparison with our observed values $\delta_1 = 0.78\pm0.02$ and $\delta_2 = 1.44 \pm
0.04$ for different models (all in the framework of a spherical relativistic fireball). $\nui$ and $\nuc$ are the
standard injection and cooling frequencies (defined e.g. in Sari \etal \cite{SPN98}). $\nuo$ spans the whole observed
band, from optical ($\sim10^{14}$~Hz) to \Xray{} ($\sim10^{18}$~Hz).}
\centering\begin{tabular}{|llllll|}\hline
Frequencies       & Ambient  & Electron     & Predicted      & \multicolumn{2}{l|}{Discrepance}  \\
order             & density  & index $p$    & slope $\delta$ & $\delta_1$  & $\delta_2$  \\ \hline
$\nuc<\nui<\nuo$  & uniform  & 1.6$\pm$0.1  & 0.92$\pm$0.02  & 7.5$\sigma$ & 13$\sigma$  \\
                  & wind     & 1.6$\pm$0.1  & 0.95$\pm$0.01  & 9.0$\sigma$ & 12$\sigma$  \\
$\nui<\nuo<\nuc$  & uniform  & 2.6$\pm$0.1  & 1.2$\pm$0.1    & 4.3$\sigma$ & 2.4$\sigma$ \\
                  & wind     & 2.6$\pm$0.1  & 1.7$\pm$0.1    & 9.3$\sigma$ & 2.6$\sigma$ \\
$\nui<\nuc<\nuo$  & uniform  & 1.6$\pm$0.1  & 0.92$\pm$0.02  & 7.5$\sigma$ & 13$\sigma$  \\
                  & wind     & 1.6$\pm$0.1  & 0.95$\pm$0.01  & 9.0$\sigma$ & 12$\sigma$  \\ \hline
\end{tabular}\label{tb:alphadelta}
\end{table}

To extract further information on the physical parameters of the explosion, we attempted to model the broad-band
spectral energy distribution (SED) of the afterglow, including the radio detection (Frail \& Berger \cite{FB02}),
millimeter upper limits (Bremer \& Castro--Tirado \cite{BCT02}; Bertoldi \etal \cite{Be02}), and \Xray{} data
(Butler \etal \cite{Bu03}). The extrapolation of the optical spectrum lies well below the value observed in the
\Xray{} band, which has moreover a harder spectrum (Fig.~\ref{fg:SED}, dashed line). This could indicate that such
emission constitutes a different component (due e.g. to the Compton process, such as in GRB\,000926; Harrison \etal
\cite{Ha01}). Alternatively, additional extinction could be responsible for the optical/\Xray{} mismatch; we therefore
fitted again all optical/NIR datasets with a powerlaw allowing for additional absorption at $z = 1.254$, adopting the
extinction law of Cardelli, Clayton \& Mathis (\cite{CCM89}). The best fit yields $\alpha = 0.85 \pm 0.07$ with
$A_V(\mathrm{host}) = 0.12 \pm 0.04$, a small but non-negligible amount. We get $\chi^2 = 51$ for 14 degrees of
freedom. The statistical improvement (with the respect to the case of no additional extinction) is admittedly low; however,
in this case the optical spectrum becomes harder, and a single component can account for both the NIR/optical and
\Xray{} emissions (see Fig.~\ref{fg:SED}, shaded region). Moreover, the spectral indices in the two bands are
strikingly similar: $\alpha_\mathrm{X} = 0.85 \pm 0.04$, $\alpha_\mathrm{opt} =
0.85\pm0.07$. Last, we note that (independently on any extinction) the decay slopes are very well constrained, and yet
in full agreement: $\delta_\mathrm{X} = 1.38\pm0.06$, $\delta_\mathrm{opt} = 1.44\pm0.04$. This seems to be a fine
tuning if the two components have a different origin.
Using \Chandra{} data, Butler \etal (\cite{Bu03}) have reported no excess column density with respect to the
Galactic value.  However, our $A_V$ corresponds to a modest $\NH = 1.95 \times 10^{20}$~cm$^{-2}$ (assuming a Galactic
gas to dust ratio), a very small column to detect at a redshift of $z = 1.254$ (the observed value is reduced by a
factor $\sim(1+z)^3 \sim 10$). In the following, we assume that indeed NIR/optical and \Xray{} emissions constitute a
single component.

Since $\alpha_\mathrm{X} < 1$, the spectrum is hard, with its peak frequency (in $\nu F_\nu$) lying above the \Xray{}
band. If the emission we see is synchrotron by a powerlaw distribution of electrons ($N(\gamma) \propto \gamma^{-p})$,
there are two possibilities, depending on whether the emitting particles are cooling rapidly or not. In the first case,
$p = 2\alpha = 1.6\pm0.1$ (`flat' distribution), while in the second case $p = 2\alpha + 1 = 2.6\pm0.1$ (a more
conventional value). Theoretical models predict several relations between the temporal and spectral slopes, depending on
a number of assumptions such as the external density profile, cooling regime, and dynamical conditions. Before the
achromatic break, however, the fireball should always follow a relativistic spherically symmetric evolution, and the
predictions are robust and easy to check (e.g. Panaitescu \& Kumar \cite{PK00}; Dai \& Cheng \cite{DC01}).
The only delicate issue regards the flat distribution case (when $p < 2$), where the evolution of the spectrum is
sensitive to the high-energy cutoff of the electron distribution, the behaviour of which is not well understood; we adopt
the prescription given e.g. by Moderski et al. (\cite{MSB00}) and Dai \& Cheng (\cite{DC01}).
Table~\ref{tb:alphadelta} shows all possible cases, presenting different ambient media and spectral shapes.  None
of them can account for the low value $\delta_1 = 0.78\pm0.02$. To eliminate such mismatch, a possible explanation
is to invoke a strong Compton cooling (either on self-synchrotron or external photons), whose emission is confined at
high energies (\mbox{$> 10$ keV}). In this case, the cooling frequency should decrease slower with time, or even increase, and
so the decay would be flatter above this frequency. Alternatively, a refreshed shock model (Sari \& \mes{} \cite{SM00})
could also alleviate the discrepance, as recently proposed by Bj\"ornsson \etal (\cite{Bj02}) for GRB\,010222.

%

\begin{acknowledgements}
We thank \"Umit Kizilo\u{g}lu for communicating to us his refined $R$-band measurement, Jens Hjorth, Leslie Hunt and Sergio
Campana for useful discussion. We also thank Scott Barthelmy for maintaining the GCN system. FT and DM acknowledge the
Italian MIUR for financial support.
\end{acknowledgements}

\end{document}